\documentclass{ws-ijmpd}

\usepackage{amsmath,graphicx,mathrsfs}

\newcommand{\beq}{\begin{equation}}
\newcommand{\eeq}{\end{equation}}
\newcommand{\eref}[1]{eq.~(\ref{#1})}

\newcommand{\leftexp}[2]{{\vphantom{#2}}^{#1}{#2}}

\begin{document}

\markboth{Dan Giang and Charles C. Dyer}
{Velocity Dominated Singularities in the Cheese Slice Universe}

%
\catchline{}{}{}{}{}
%

\title{VELOCITY DOMINATED SINGULARITIES IN THE CHEESE SLICE UNIVERSE}

\author{DAN GIANG
}

\address{Department of Physics, University of Toronto, 60 St. George St.\\
Toronto, Ontario M5S 1A7,
Canada
\\
dgiang@physics.utoronto.ca}

\author{CHARLES C. DYER}

\address{Department of Astronomy, University of Toronto, 50 St. George St.\\
Toronto, Ontario M5S 3H4, Canada\\
dyer@astro.utoronto.ca}

\maketitle


\begin{abstract}
We investigate the properties of spacetimes resulting
from matching together exact solutions using the Darmois matching
conditions.  In particular we focus on the asymptotically velocity
term dominated property (AVTD).  We propose a criterion that can be
used to test if a spacetime constructed from a matching can be
considered AVTD.
Using the Cheese Slice universe as an example, we
show that a spacetime constructed 
from a such a matching can inherit the AVTD property from the original
spacetimes.  Furthermore the
singularity resulting from this particular matching is an AVTD
singularity.
\end{abstract}


\section{Introduction}

The Friedmann-Lema\^{i}tre-Robertson-Walker (FLRW)
\cite{friedmann,lemaitre,robertson,walker} spacetime is a good
cosmological model with which to approximate our universe, but it does
not account for inhomogeneities that are observed in small and large scale
structure.  In an attempt to model more realistic cosmologies
that can account for large scale inhomogeneities there have been some
models proposed that are constructed my matching together various
solutions.  The most famous of which is the Einstein-Straus ``Swiss
Cheese'' models \cite{einstein-straus}.  More recently there has been
a planar model proposed by matching together FLRW and Kasner
\cite{kasner} spacetimes.  In relation to the Einstein-Straus models,
these models have been termed the ``Cheese Slices'' universe.  Such
constructions could be of interest in relation to the observed
layering in the distribution of galaxies reported by Broadhurst et
al. \cite{broadhurst}.

However, if such cosmologies are to be used as
models of our universe, we must understand their implications at all
stages of evolution including at the initial singularity.  In
particular, when a cosmology is constructed by matching together
different spacetimes, what structure does the singularity inherit from
the matching?  Does a matching at late times naturally imply, in any
sense, a ``well behaved'' matching at the singularity?  Such
questions are difficult to formulate in the most general terms, thus
we choose to focus on the aspect of velocity dominated singularities.

Belinskii, Khalatnikov and Lifshitz (BKL) \cite{bkl1970} have
conjectured that in a generic singularity the evolution towards the
singularity is independent of spatial curvature.  Since then other
authors have attempted to formulate more precise definitions of this
property such as Eardley, Liang and Sachs \cite{els} in their
definition of velocity dominated singularities.  Their definition has
been generalized by Isenberg and Moncrief \cite{moncrief} and is known
as the asymptotically velocity term dominated (AVTD) property.

In
particular with the Cheese Slice universe, we have a matching of two
AVTD spacetimes with
asymptotically velocity term dominated singularities (AVTDS).  The
question arises of whether or not the singularity in the Cheese
slice universe inherits the AVTD property from the spacetimes used in
its construction.  In section \ref{AVTD} we show how a spacetime
constructed from a matching can be 
considered to be AVTD.  We will also detail the Darmois
matching conditions \cite{darmois} that we adopt as our matching criteria
throughout.  Then in section \ref{cheese} we show that the Cheese
Slice singularity is indeed an AVTDS.

In the following Greek indexes indicate four dimensions,
$\alpha,\beta,\gamma\ldots=\{0,1,2,3\}$.  Latin indexes indicate
three dimensions, $a,b,c\ldots=\{1,2,3\}$ and upper case indexes
indicate two dimensions, $A,B,C\ldots=\{1,2\}$.  We will also refer to
spacetimes constructed by matching together different solutions as a
``matched spacetime''.

\section{The AVTD Property of a Matched Spacetime}\label{AVTD}

\subsection{Definitions}

Let $U$ be a spacetime with metric $g_{\alpha\beta}$ and coordinates
$x^\alpha$.  We 
begin by choosing a spatial foliation with intrinsic coordinates
$\xi^a$ on each leaf of the foliation.  Next we identify  
the intrinsic metric, 
\beq\label{gdef}
\gamma_{ab} = \frac{\partial x^\alpha}{\partial \xi^a}
              \frac{\partial x^\beta}{\partial \xi^b} g_{\alpha\beta}, 
\eeq
and extrinsic curvature, 
\beq\label{Kdef}
K_{ab} = \frac{\partial x^\alpha}{\partial \xi^a}
         \frac{\partial x^\beta}{\partial \xi^b} \nabla_\alpha n_\beta, 
\eeq
of the spacelike three surfaces, where $n_\alpha$ is the normal to the
surface.  The mean 
curvature is then $K=K^a_a$.  The timelike foliation vector,
$\partial/\partial t$, where $t$ is a timelike coordinate that labels 
successive leaves of the foliation, describes the evolution of the
three surface and is related to the surface normal via the lapse $N$
and shift $M_\alpha$,
\beq
\frac{\partial}{\partial t} = N n_\alpha + M_\alpha.
\eeq
The matter density, $\rho$, momentum, $J_a$, and spatial stress
densities, $S_{ab}$, must also be considered.  These quantities must
satisfy the Einstein Field Equations written in the form of constraint
equations \cite{moncrief}
\beq\label{E1}
\leftexp{(3)}{R} - K^{ab}K_{ab} + K^2 = 2\rho,
\eeq
\beq\label{E2}
\leftexp{(3)}{\nabla}_a K^a_b - \leftexp{(3)}{\nabla}_b K = - J_b
\eeq
and evolution equations,
\beq\label{E3}
\frac{\partial}{\partial t}\gamma_{ab} =
 -2NK_{ab} + \mathscr{L}_M \gamma_{ab},
\eeq
\beq\label{E4}
\frac{\partial}{\partial t}K^a_b =
 N\left[ \leftexp{(3)}{R}^a_b + KK^a_b + 
         S^a_b + \frac 1 2 \gamma^a_b\left(\rho-S^c_c\right)\right ]
 - \leftexp{(3)}{\nabla}^a\leftexp{(3)}{\nabla}_b N + \mathscr{L}_M K^a_b,
\eeq
where $\leftexp{(3)}{R}$ and $\leftexp{(3)}{R}_{ab}$ are the spatial
Ricci scalar and Ricci tensor respectively.  $\leftexp{(3)}{\nabla}$
is the three 
dimensional covariant derivative and $\mathscr{L}_M$ is the Lie
derivative in the direction of $M_\alpha$.  Also geometrized units
have been used where $8\pi G=1$

Next the velocity term dominated solutions (VTD) are defined by
neglecting all the spatial derivatives in the field equations.  This
leads to the VTD constraint equations \cite{moncrief},
\beq\label{V1}
\tilde K^{ab}\tilde K_{ab} + \tilde K^2 = 2\tilde\rho,
\eeq
\beq\label{V2}
\leftexp{(3)}{\nabla}_a \tilde K^a_b - 
\leftexp{(3)}{\nabla}_b \tilde K = - \tilde J_b, 
\eeq
and the VTD evolution equations,
\beq\label{V3}
\frac{\partial}{\partial t}\tilde \gamma_{ab} =
 -2N\tilde K_{ab},
\eeq
\beq\label{V4}
\frac{\partial}{\partial t}\tilde K^a_b =
 N\left[ \tilde K\tilde K^a_b + \tilde S^a_b
         + \frac 1 2 \tilde \gamma^a_b\left(\rho-\tilde S^c_c\right)
  \right ].
\eeq
Note that in general the spatial derivatives in $\rho$, $J_a$ and
$S_{ab}$ are removed as well.  We use the $\tilde{}$ to indicate their
distinctiveness from the quantities in the Einstein
eq.~(\ref{E1})--(\ref{E4}). 

Solutions of the field eq.~(\ref{E1})--(\ref{E4}) are then defined
to be AVTD if in the limit of large $t$ they approach the solutions to
the VTD 
eq.~(\ref{V1})--(\ref{V4}).  That is, as $t\rightarrow\infty$, the
values of 
\beq\label{AVTDcondition}
\left\{\gamma_{ab}, K_{ab}, \rho, J_a, S_{ab}\right\} -
\left\{\tilde \gamma_{ab}, \tilde K_{ab}, \tilde \rho, \tilde J_a, \tilde
S_{ab}\right\}=0.
\eeq
A singularity is said to be an 
AVTDS if the spacetime is AVTD and the foliation is chosen such that
the singularity is approached as $t\rightarrow\infty$.

\subsection{The Matching of Spacetimes}

We will use the Darmois matching conditions \cite{darmois} to
piece together different spacetimes.  Suppose we have two regions of spacetime,
$V^-$ and $V^+$ with metrics $g^-_{\alpha\beta}$ and
$g^+_{\alpha\beta}$ respectively. 
According to the Darmois conditions, these two 
regions of spacetime match across a hypersurface $\Sigma$ if the first
and second fundamental forms, calculated in terms of the coordinates
on $\Sigma$, are identical.  More precisely, let $\chi^a$ be the
coordinates on $\Sigma$.  The first and second fundamental forms are
defined as,
\beq\label{ff}
\Upsilon_{ab} = \frac{\partial x^\alpha}{\partial \chi^a}
              \frac{\partial x^\beta}{\partial \chi^b} g_{\alpha\beta}, 
\eeq
and
\beq\label{sf}
\Omega_{ab} = \frac{\partial x^\alpha}{\partial \chi^a}
         \frac{\partial x^\beta}{\partial \chi^b}\nabla_\alpha n^\Sigma_\beta, 
\eeq
where $n^\Sigma_\alpha$ is the normal to $\Sigma$.  This is identical
in form to eq.~(\ref{gdef}) and (\ref{Kdef}), however, we state the
definition and notation here
to emphasize the distinction between the timelike surface $\Sigma$ and the
spatial three-surfaces of the foliation.
The Darmois conditions are then
\beq\label{d1}
\Upsilon^-_{ab}=\Upsilon^+_{ab}
\eeq
and
\beq\label{d2}
\Omega^-_{ab}=\Omega^+_{ab}.
\eeq
The superscript $-$ and $+$ indicate the quantity is calculated from
either $V^-$ or $V^+$ with the appropriate metrics and normals.  If
the Darmois conditions (\ref{d1}) and (\ref{d2}) are satisfied then we
can match $V^-$ and $V^+$ along $\Sigma$ resulting in a new exact
solution of the Einstein Field Equations with no additional stress
energy required along $\Sigma$.

We now describe in what sense a matched spacetime could be considered
AVTD.  Let $W$ be the spacetime 
constructed from the matching of $V^-$ and $V^+$ across the
 surface $\Sigma$.  Also, let $\Pi^\pm_{t^\pm}$ denote leaves
of a foliation of $V^\pm$, parametrized by $t^\pm$, such that $V^\pm$
is AVTD.  In general $t^-$ and $t^+$ are different time coordinates.
If each leaf of the foliation $\Pi^-_{t^-}$ matches with each leaf
of the foliation $\Pi^+_{t^+}$ along the surface $\Sigma$, then this
constitutes a foliation of $W$ such that $W$ is AVTD.

Note that the corresponding VTD solutions must also match across the
surface $\Sigma$ in the same manner.

\begin{figure}
\begin{center}
\includegraphics[height=5.6666cm, width=7.8cm]{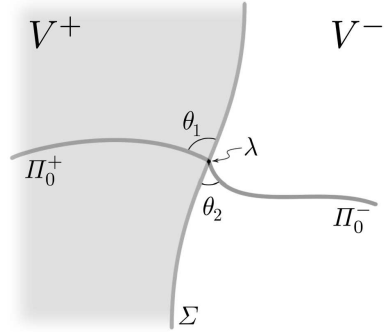}
\caption{A leaf of the foliation on each side is singled out,
  $\Pi_0^\pm$.  The intersection of $\Pi_0^\pm$ with $\Sigma$ is the
  what we refer to as the corner, $\Lambda$, which is itself a
  two-surface.}
\label{corner}
\end{center}
\end{figure}

To clarify the matching of the $\Pi^-_{t^-}$ with  $\Pi^+_{t^+}$, let
us single out one leaf of the foliation on each side and call them
$\Pi^\pm_0$.  See Figure~\ref{corner}.
$\Pi^\pm_0$ are spatial three-surfaces in $V^\pm$.  We wish to match
$\Pi^-_0$ with $\Pi^+_0$ across the surface $\Sigma$.  However,
$\Sigma$ is a timelike three-surface and the intersection of $\Pi^\pm$
with $\Sigma$ is a spatial two-surface.  Let us call this two-surface
the ``corner'' and denote it by $\Lambda$ with the coordinates
$\zeta^A$.
There is also a two-dimensional space of normals to $\Lambda$.  Let
$m^A_\alpha$ be an orthonormal basis for this space. 

Fortunately the matching
conditions at a corner have already been thoroughly investigated by
Taylor \cite{taylor}.  If the Darmois matching conditions are
satisfied along two intersecting hypersurfaces then certain conditions
must be true at the corner.  Thus the matching conditions at a
corner are derived from the Darmois conditions.
We quote the result here.

The first and second fundamental forms on the corner are defined as
\beq
\hat\gamma_{AB} = \frac{\partial x^\alpha}{\partial \zeta^A}
              \frac{\partial x^\beta}{\partial \zeta^B} g_{\alpha\beta}, 
\eeq
\beq
\hat K^C_{AB} = \frac{\partial x^\alpha}{\partial \zeta^A}
         \frac{\partial x^\beta}{\partial \zeta^B} \nabla_\alpha m^C_\beta. 
\eeq
There is also a torsion vector defined as,
\beq
\tau_A = \frac{\partial x^\alpha}{\partial \zeta^A}
         m^{1\beta}\nabla_\alpha m^2_\beta.
\eeq
Finally, let $\theta_A^\pm$ denote the angle between $\Pi^\pm$ and
$\Sigma$.

Then the two three-spaces $\Pi^-$ and $\Pi^+$ match at the corner
$\Lambda$ if
\beq\label{corner1}
\hat\gamma^-_{AB} =\hat\gamma^+_{AB},
\eeq
\beq\label{corner2}
\hat K^{C-}_{AB} = \hat K^{C+}_{AB},
\eeq
\beq\label{corner3}
\tau^-_A = \tau^+_A
\eeq
and
\beq\label{corner4}
\theta_A^- = \theta_A^+,
\eeq
where, as above, the superscripts $-$ and $+$ indicates that the
quantity is calculated in $V^-$ or $V^+$.

\section{Singularities in the Cheese Slice Universe}\label{cheese}

We now turn to an example of a spacetime constructed from a matching
of exact solutions that satisfy the Darmois conditions and
exhibits the AVTD property in the terms described above.
The Cheese Slice universe is constructed by matching together FLRW and
Kasner spacetimes along planar surfaces \cite{landry}.  Note that the
matched spacetime does not require any additional stress-energy on the
matching surface.  Such a spacetime could be used to model large scale
inhomogeneities in our universe.  The ability to combine FLRW
regions with large vacuum regions makes the Cheese Slices a more
comprehensive cosmological model than the FLRW on its own.

\subsection{The Matchings}

\begin{figure}
\begin{center}
\includegraphics[height=5.0cm, width=11cm]{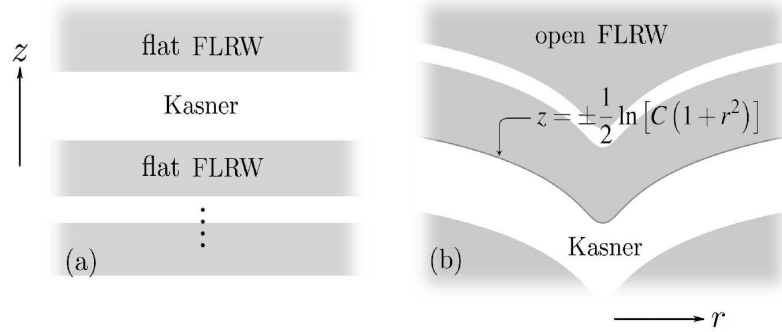}
\caption{(a) The Cheese slice universe constructed by matching together
  flat FLRW and Kasner spacetimes along the surface $z=const$.  (b)
  The Cheese Slice universe constructed with open FLRW regions using a
  different matching surface.  Both matchings can be carried on
  indefinitely with layers of arbitrary thicknesses.} 
\label{slices}
\end{center}
\end{figure}

The FLRW line element in cylindrical coordinates is given by,
\beq\label{fmetric}
ds^2_F = -dt^2 + a^2(t)\left[ \frac{dr^2}{1-kr^2} + r^2d\phi^2 +
(1-kr^2)dz^2 \right],
\eeq
with $k=\{-1, 0, 1\}$.  The Kasner line element is given by,
\beq\label{kmetric}
ds^2_K = -dT^2 + T^{2p_1}dX^2 + T^{2p_2}dY^2  + T^{2p_3}dZ^2,
\eeq
with the restrictions 
\beq
p_1+p_2+p_3=1=p_1^2+p_2^2+p_3^2.
\eeq
If the FLRW spacetime is flat and pressure free ($k=0$,$a(t)=t^{4/3}$)
and the Kasner exponents are $p_1=p_2=\frac 2 3$, $p_3=-\frac 1 3$
then one can show that the first and second fundamental forms,
eq.~(\ref{ff}) and (\ref{sf}), are
identical when calculated on the surface $z=z_0$ in the FLRW
spacetime and $Z=Z_0$ in the Kasner spacetime, where $z_0$ and $Z_0$
are constants.  Thus the Darmois
conditions are satisfied and we can match
these two spacetimes along this surface to construct the
Cheese Slice universe.  Note that this matching can be repeated
indefinitely and with layers of arbitrary thicknesses.  See
Figure~\ref{slices}(a).

A similar matching can also take place with an open FLRW universe
($k=-1$), but the matching surface must take a different form
\cite{oliwa}.  In this case the surface must be of the form 
\beq
z=\pm\frac 1 2 \ln[C(1+r^2)],
\eeq
where $C$ is a positive constant.  Refer to Figure~\ref{slices}(b) for
an illustration of this matching surface.

\subsection{The Singularities}

Both the Kasner and the FLRW spacetimes have an initial singularities
($t=0=T$).  We look at
the cases of the flat FLRW matching and the 
open FLRW matching in turn. 

\subsubsection{Case (i) Flat FLRW, $k=0$}

The coordinates defined in
eq.~(\ref{fmetric}) and (\ref{kmetric}) single out a natural
foliation that we will use to check the AVTD property for the flat case.
The Kasner spacetimes satisfies the VTD
eq.~(\ref{V1})--(\ref{V4}) directly therefore it is trivially AVTD.
With the pressure free FLRW spacetime the flat case satisfies the VTD
eq.~(\ref{V1})--(\ref{V4}) as well with the following quantities,
\begin{eqnarray}
&\tilde\gamma_{ab}= a^2 \, \textrm{diag}\left(1,r^2,1\right), \qquad
\tilde K_{ab} = a_{,t} a \, \textrm{diag}\left(1,r^2,1\right),&\nonumber
\\
&\tilde \rho = 3\left(\frac{a_{,t}}{a}\right)^2, 
\quad 
\textrm{and}
\qquad\tilde J_a = 0 = \tilde S_{ab},&
\end{eqnarray}
where $\phantom{a}_{,t}=\frac{\partial}{\partial t}$.  Thus both sides
are AVTD.  Furthermore, we
can make the coordinate transformation $\tau=-\ln t$ to set the
singularity at $\tau=\infty$ and all the conditions of an AVTDS are
satisfied.

To show that the matched spacetime is also AVTD with the chosen
foliation we must check that the corner conditions,
eq.~(\ref{corner1})--(\ref{corner4}), are satisfied.

On the FLRW side the corner is defined as $z=z_0$ and $t=t_0$, with
$z_0$ and $t_0$ being constants.
Thus we have,
\begin{eqnarray}
&\hat\gamma^+_{AB}=a^2\,\textrm{diag}\left(1,r^2\right),
\qquad
\hat K^{+1}_{AB}=-a_{,t} a \,\textrm{diag}\left(1,r^2\right),
&\nonumber\\
&\hat K^{+2}_{AB}=0
\qquad
\textrm{and}
\qquad
\tau^{+}_{A}=0.&
\end{eqnarray}
On the Kasner side the corner is defined as $Z=Z_0$ and $T=T_0$, with
$Z_0$ and $T_0$ being constants.
Thus we have,
\begin{eqnarray}
&\hat\gamma^-_{AB}=\textrm{diag}\left(T^{2p_1},T^{2p_2}\right),
\qquad
\hat K^{-1}_{AB}=T^{-1}\,
                \textrm{diag}\left(-p_1T^{2p_1},-p_2T^{2p_2}\right), 
&\nonumber\\
&\hat K^{-2}_{AB}=0
\qquad
\textrm{and}
\qquad
\tau^{-}_{A}=0.&
\end{eqnarray}
If we choose the coordinates on the corner as $\zeta^A=\{u,v\}$,
parametrize the surface as $r \cos\phi=u=X$ and $r \sin\phi=v=Y$ we
can satisfy eq.~(\ref{corner1})--(\ref{corner3}).
Recall that $a=t^{2/3}$ and $p_1=p_2=2/3$.
Furthermore the surfaces defining the corners are orthogonal on both
sides and the matching surface subtends an angle of $\pi$ as seen
from either side and thus
eq.~(\ref{corner4}) is also satisfied.  Therefore we have a
matching at the corner and the flat Cheese Slice universe is AVTD.

Also, notice that for the matching to take place we have also
identified the time coordinates $t=T$.  With the coordinate
transformation $\tau=-\ln t=-\ln T$ we can set the singularity at
$\tau=\infty$ and the conditions for an AVTDS are satisfied.

\subsubsection{Case (ii) Open FLRW, $k=-1$}

In general the AVTD property is highly dependent on the choice of
foliation.  A spacetime that is AVTD in one foliation might not appear
to be
AVTD in another, thus we must be careful in our choice of foliation.
To show that the open Cheese Slices can be AVTD we make the following
transformation on the FLRW side,
\beq
\tilde z = z-\frac 1 2 \ln(1+r^2).
\eeq
The FLRW metric (\ref{fmetric}) then becomes,
\beq
ds^2_F = -dt^2 + a^2(t)\left[dr^2 + r^2d\phi^2 + 2rdrd\tilde z + (1+r^2)d\tilde z^2\right].
\eeq
On the Kasner side we will make the transformations,
\beq
R  = \sqrt{X^2 + Y^2},
\eeq
\beq
\Phi = \arctan(Y/X),
\eeq
\beq
\tilde Z = Z - \frac{9}{16b^5} 
           \left[-3bT^{\frac 1 3}\sqrt{1+b^2T^{\frac 2 3}}
                 +2b^3T\sqrt{1+b^2T^{\frac 2 3}}
                 +3\ln\left(\sqrt{1+b^2T^{\frac 2 3}}
                            +bT^{\frac 1 3}\right)\right]
\eeq
and
\beq
\tilde t = \frac{3}{2b^3}\left[bT^{\frac 1 3}\sqrt{1+b^2T^{\frac 2 3}}
                           -\ln\left(\sqrt{1+b^2T^{\frac 2 3}}
                                    +bT^{\frac 1 3} \right)
		          \right],
\eeq
where $b$ is a positive constant. 
With these transformations the Kasner metric (\ref{kmetric}) becomes,
\beq
ds^2_K = -d\tilde t^2 + T^{\frac 4 3}\left(dR^2+R^2d\Phi^2\right)
                + 2bd\tilde td\tilde Z
                + T^{-\frac 2 3}d\tilde Z^2.
\eeq
The matching now takes place along the surface $\tilde z=\tilde z_0$
on the FLRW 
side and $\tilde Z=\tilde Z_0$ on the Kasner side, with $\tilde z_0$
and $\tilde Z_0$ being 
constants.  The coordinates, $\phi=\Phi$ and $t=\tilde t$, can be 
identified along the matching surface.  We must also have $r=\frac23bR$ and
$a^2(t)=\frac{9}{4b^2}T^{\frac 4 3}(\tilde t)$.

We will use this new foliation to check the AVTD property.  Starting
with the FLRW case it is straightforward to check that
\eref{E1}--(\ref{E4}) are satisfied with the following quantities,
\beq
\begin{array}{l}
 \gamma_{11}= a^2, \\
 \gamma_{13}= a^2r, \\
 \gamma_{22}= a^2r^2, \\
 \gamma_{33}= a^2(r^2+1),
 \end{array}
\qquad
\begin{array}{l}
 K_{11}= a_{,t}a, \\ 
 K_{13}= a_{,t}ar, \\ 
 K_{22}= a_{,t}ar^2, \\ 
 K_{33}= a_{,t}a(r^2+1),\nonumber
\end{array}
\eeq
\beq
\rho = 3\left(\frac{a_{,t}^2-1}{a^2}\right), 
\quad 
\textrm{and}
\qquad J_a = 0 = S_{ab}.
\eeq
The corresponding VTD solution is the flat FLRW solution.  We can see
that \eref{AVTDcondition} is satisfied and thus the open FLRW is AVTD.

Turning to the Kasner case we find that it also satisfies the VTD
eq.~(\ref{V1})--(\ref{V4}) with the lapse and shift being,
\beq
N = \sqrt{1+b^2T^{\frac 2 3}} \qquad\textrm{and}\qquad M_a=(0,0,b)
\eeq
respectively.  Therefore it is once again trivially AVTD.

Next we check the corner conditions,
eq.~(\ref{corner1})--(\ref{corner4}).  The corners on the FLRW
and 
Kasner sides are defined as $\{\tilde z=\tilde z_0,t=t_0\}$ and
$\{\tilde Z=\tilde Z_0,\tilde t=\tilde t_0\}$ respectively with $t_0$ and
$\tilde t_0$ being 
constants.  Recall that the coordinates are such that $r=\frac23bR$ and
$\Phi=\phi$.  Let us use the superscripts $-$ to denote the Kasner side
and $+$ to denote the FLRW side.  The first corner condition,
\eref{corner1}, is satisfied with,
\beq
\hat\gamma^-_{AB}
=\textrm{diag}(\frac{9}{4b^2}T_0^{\frac 4 3},RT_0^{\frac 4 3})
=\textrm{diag}(a_0^2,ra_0^2)
=\hat\gamma^+_{AB},
\eeq
where $T_0=T(\tilde t_0)$ and $a_0=a(t_0)$.  Let an orthonormal basis
of the corner be chosen on both sides such that, 
\beq
\begin{array}{ll}
m^{-1}_\alpha=(0,0,0,T_0^{-\frac 1 3}\sqrt{1+b^2T_0^{\frac 2 3}})\qquad
&  m^{+1}_\alpha=(0,0,0,a_0)\\
m^{-2}_\alpha=(1,0,0,-b)
&  m^{+2}_\alpha=(1,0,0,0)
\end{array}.
\eeq
Then the second corner condition, \eref{corner2}, is satisfied with,
\beq
\hat K^{-1}_{AB} 
= \frac 3 2 bT_0^{\frac23}\,\textrm{diag}(1,\frac{4}{9}b^2R^2)
= a_0\,\textrm{diag}(1,r^2)
= \hat K^{+1}_{AB}
\eeq
and
\beq
\hat K^{-2}_{AB} 
= \frac 2 {3b^2} T_0^{\frac13}\sqrt{1+b^2T_0^{\frac23}}\,
  \textrm{diag}(1,\frac{4}{9}b^2R^2)
= a_{,t0}a_0\,\textrm{diag}(1,r^2)
= \hat K^{+2}_{AB}.
\eeq
The torsion is identically zero on both sides satisfying
\eref{corner3}.
On the FLRW side the foliation is orthogonal to the
matching surface and the matching surface itself subtends an angle of
$\pi$ about the corner.
On the Kasner side, the foliation is not orthogonal to the matching
surface.
Fortunately the matching surface also subtends
an angle of $\pi$ about the corner.  This ensures condition
\eref{corner4} is satisfied on both sides.

Similar to the flat matching, the time coordinate may be
transformed as desired, since it is identical on both sides, to ensure
that the singularity is reached as $t\rightarrow\infty$ and the
singularity may be considered an AVTDS.

\begin{figure}
\begin{center}
\includegraphics[height=5.5cm, width=9cm]{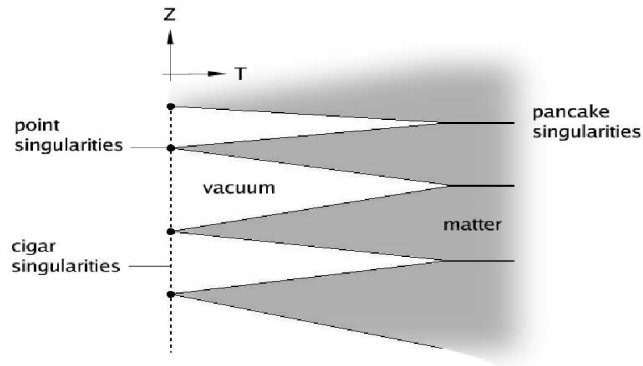}
\caption{The Singularities of the Cheese Slice Model.  The initial
  singularity is a chain of cigar singularities, corresponding to the
  Kasner vacuum regions, and point singularities, corresponding to
  FLRW regions.  At late time, the vacuum regions become arbitrarily
  thin pancake-like singularities.}
\label{fig_final}
\end{center}
\end{figure}

Finally, let us illustrate how this singularity in the Cheese Slice
universe manifests itself.  In the Kasner regions the initial
singularity is of a cigar type and at late times
the Kasner regions have pancake-like singularities.
In the FLRW slices we have an initial point-like singularity and no
singularities at late times.  Thus we can visualize the initial
singularity of the Cheese Slices as an inhomogeneous chain of
cigar-like singularities joined by point-like singularities.  At late
times, the Cheese Slices become an inhomogeneous matter filled space
with pancake-like singularities throughout, as illustrated in Figure (\ref{fig_final}).

\section{Conclusions}

We have proposed a criterion with which we may consider a matched
spacetime to be AVTD.  First, both sides of the matched
spacetime must be AVTD.  Secondly each leaf of the chosen foliation
must also match across the surface at an intersection that we refer to
as the corner.  We have also demonstrated this with the example of
the Cheese Slice universe.  The flat Cheese Slice satisfies these
conditions in a straightforward manner whereas the open Cheese slice
required more effort to find a foliation that satisfied the AVTD
property and the matching conditions.  In a general matching it may be
difficult to find a foliation that is consistent with the matching and
the AVTD property.  
However, as we have shown, it is
possible in the case of Cheese Slice universe for the
singularity to inherit the AVTD property from the different spacetimes
used in its construction.

In addition to modeling inhomogeneities, these models of matched
spacetimes are also very useful in investigating what
matching conditions could tell us about the properties of spacetimes
themselves.  
For example, we conjecture that any spacetime that can be smoothly
matched to 
an AVTD spacetime, using the Darmois conditions, must necessarily be
AVTD.  The resulting matched spacetime would also be AVTD.  The
general proof of this remains to be seen and is open to 
investigation.  If true, this could lead the way to using the Darmois
conditions to prove AVTD properties of other spacetimes.

\section*{Acknowledgments}

    The work of DG was supported in part by a postgraduate scholarship
    from the Natural Sciences and Engineering Council of Canada.  CCD
    acknowledges the support of the National Sciences and Engineering
    Council of Canada via a Discovery Grant.  We also acknowledge
    useful comments from referees.



\begin{thebibliography}{10}

\providecommand{\url}[1]{{#1}}
\providecommand{\urlprefix}{URL }
\expandafter\ifx\csname urlstyle\endcsname\relax
  \providecommand{\doi}[1]{DOI~\discretionary{}{}{}#1}\else
  \providecommand{\doi}{DOI~\discretionary{}{}{}\begingroup
  \urlstyle{rm}\Url}\fi

\bibitem{friedmann}
A. Friedmann, {\it Z. Phys.} \textbf{10} (1922) 377

\bibitem{lemaitre}
G. Lema\^{i}tre, 
\newblock {\it Annales Soc. Sci. Brux. Ser. I
  Sci. Math. Astron. Phys.} \textbf{A53}
  (1933) 51--85 

\bibitem{robertson}
H. P. Robertson, {\it Astrophys. J.} \textbf{82} (1935) 284 

\bibitem{walker}
A. G. Walker, {\it proc. London math. Soc.} \textbf{42}(90) (1936)

\bibitem{einstein-straus}
A. Einstein and E. G. Straus, 
\newblock {\it Reviews of Modern Physics} \textbf{17} (1945) 120--124 

\bibitem{kasner}
E. Kasner, 
\newblock {\it American Journal of Mathematics} \textbf{43}(4), (1921)
217--221 

\bibitem{broadhurst}
T. J. Broadhurst, R. S. Ellis, D. C. Koo and A. S. Szalay, 
\newblock {\it Nature } \textbf{343} (1990) 726--728 

\bibitem{bkl1970}
V. Belinskii, E. Lifshitz and I. Khalatnikov, 
\newblock {\it Adv. Phys. } \textbf{19} (1970) 525--573 

\bibitem{els}
D. Eardley, E. Liang and R.  Sachs, 
\newblock {\it J. Math. Phys.} \textbf{13}(1) (1972) 99--107 

\bibitem{moncrief}
J. Isenberg and V. Moncrief,  
\newblock {\it Annals of Physics} \textbf{199} (1990) 84--122

\bibitem{darmois}
G. Darmois, {\it M\'emorial des Sciences Mathematiques} \textbf{25} (1927) 25

\bibitem{taylor}
J. P. W. Taylor, 
\newblock {\it Class. Quantum Grav.} \textbf{21} (2004) 3705--3715

\bibitem{landry}
C. C. Dyer, S. Landry and E. B. Shaver, 
\newblock {\it Physical Review D} \textbf{47}(4) (1993) 1404--1406

\bibitem{oliwa}
C. C. Dyer and C. Oliwa, 
\newblock {\it Class. Quantum Grav.} \textbf{18} (2001) 2719--2729

\end{thebibliography}
\end{document}